\documentclass{article}
\usepackage{spconf,amsmath,graphicx,tikz,calc,enumitem,floatrow,subcaption,algorithm,algpseudocode,numprint,comment}
\usepackage[keeplastbox]{flushend}
\usetikzlibrary{calc}

\usepackage[hyphens,spaces,obeyspaces]{url}
\usepackage[breaklinks]{hyperref}
\hypersetup{breaklinks=true}
\usepackage{cite}
\usepackage{url}
\usepackage{ragged2e}
\usepackage{amssymb}

\usepackage{flushend} % balanced references
\usepackage{booktabs}
\usepackage{cleveref}
\crefname{figure}{Fig.}{}
\crefname{table}{Table}{}
\crefname{section}{Section}{}
\crefname{equation}{}{}

\newcommand\rightlast{\leftskip0ptplus1fil\rightskip0ptplus-1fil\parfillskip0ptplus1fil}\DeclareCaptionJustification{rightlast}{\rightlast}

\ninept

\npfourdigitnosep % no separator for 4-digit numbers, per IEEE style guide

\algrenewcommand\textproc{}

\newcommand{\hg}[1]{{ {\textcolor{blue}{[HG: #1]}}}}
\newcommand{\vv}[1]{{ {\textcolor{green}{[VV: #1]}}}}
\newcommand{\zf}[1]{{ {\textcolor{teal}{[ZF: #1]}}}}

\newcommand{\xs}{\mathbf{x}_0}
\newcommand{\obj}{\mathcal{O}}
\newcommand{\rot}{\mathcal{R}}

\def\x{{\mathbf x}}

\title{Fast Acoustic Scattering using Convolutional Neural Networks}

\name{Ziqi Fan$^1$\sthanks{Work done as research intern at Microsoft Research, Redmond}, Vibhav Vineet$^2$, Hannes Gamper$^2$, Nikunj Raghuvanshi$^2$}
\address{$^1$University of Florida, Gainesville\\
$^2$Microsoft Research, Redmond}
  
\begin{document}
\maketitle

\begin{abstract}
Diffracted scattering and occlusion are important acoustic effects in interactive auralization and noise control applications, typically requiring expensive numerical simulation. We propose training a convolutional neural network to map from a convex scatterer's cross-section to a 2D slice of the resulting spatial loudness distribution. We show that employing a full-resolution residual network for the resulting image-to-image regression problem yields spatially detailed loudness fields with a root-mean-squared error of less than 1 dB, at over 100x speedup compared to full wave simulation.
\end{abstract}


\section{Introduction}
\label{sec:intro}
\begin{comment}
\hg{Here is how I would structure the introduction:
a) Scattering / occlusion are important but difficult to model
b) Wave simulations yield answer, but are extremely costly
c) Authors have looked at real-time solutions (e.g., Facebook / Google?) - but they have limitations
d) Recently, machine learning showed promising results in many audio-related problems: speech synthesis, localization, T60 estimation, etc., indicating that ML models are capable of modelling highly complex acoustic phenomena
e) Pulkki and Perez introduced approaches for modelling scattering
f) ``Here we show...'': Treat problem as mapping directly from geometry to parametric acoustic field, train DNN to learn mapping (rather than learning the parameters of an existing physical model or using hand-crafted features)}
\end{comment}

%{Besides exterior and interior sound propagation, the noise-and-vibration behavior of a dynamic system can also be predicted using neural networks. For example, Tsokaktsidis et. al. investigated predicting noise transfer in a system composed of two coupled beams using a neural network \cite{Tsokaktsidis2019noisetransfer} with \numprint{5} neurons. Their training data was acquired through simulation using the finite element method (FEM) and a neural network with a single hidden layer with \numprint{5} neurons was used for training and prediction. In using neural networks, the capability a neural network generalizes to unseen cases is of great significance and the generalization capability is closely related to the diversity of its training data.}

Fast evaluation of wave scattering and occlusion from general object shapes is important for diverse applications, such as optimizing baffle shape in outdoor noise control \cite{bies2017engineering,beranek1992noise}, and real-time auralization in games and mixed reality \cite{vorlander:2007,Raghuvanshi:2018:Triton}.
Modeling diffraction is critical since acoustical wavelengths span everyday object sizes. Wave solvers capture diffraction and can achieve real-time execution in restricted cases\cite{savioja:2010,allen_aerophones:2015} but in general they remain quite expensive, even with hardware acceleration \cite{fan2019computation,raghuvanshi_ARD_GPU:2009}. While pre-computed wave simulation is viable for real-time auralization \cite{Raghuvanshi:2018:Triton}, it disallows arbitrary shape changes at run-time. Geometric (ray-based) approaches can handle dynamic geometry but diffraction remains challenging due to the inherent zero-wavelength approximation \cite{SaviojaGASurvey:2015}.

\vspace{2pt}

%\zf{Add my recent paper about HRTF modeling here. Based on the performance of our optimized BEM solver, real-time solver is tough.}

%Prior work exists on efficient alternatives to solve the diffraction problem in real time. Google et al... %\zf{How to complete this paragraph? My understanding for Google resonance is that it is a 3D audio rendering engine. I am not quite sure if it considers wave acoustics. More importantly, I can't find any specific detail describing how it is implemented, thus, I cannot comment on it.}

%We ask the question if deep neural networks may offer a route to significantly accelerate such evaluation.

We propose a machine learning approach for fast modeling of diffracted occlusion and scattering. Previously, machine learning has been successfully applied in acoustic signal processing problems including speech synthesis \cite{ze2013statistical,wu2015deep}, source localization \cite{he2018deep,ferguson2018sound}, blind estimation of room acoustic parameters from reverberated speech \cite{eaton2016estimation,Genovese2019}, binaural spatialization \cite{Tenenbaum2019auralization}, and structural vibration \cite{Tsokaktsidis2019noisetransfer}.
{P\`erez} et al.\ \cite{falcon2018machine} used a fully-connected neural network to learn 
the effect of re-configuring the furniture layout of a single room on acoustical parameters, including reverberation time ($T_{60}$) and sound pressure level (SPL), 
%\hg{is this the same as $L$ in Eq. 2? Otherwise rename / remove}
at a few listener locations. %\zf{It is different from the loudness field in equation 2}. 
%\hg{were other parameters estimated?} 
%\zf{Yes. Should we include a list of all these parameters or just provide several examples?}
%\cite{falcon2018machine} predicted room acoustic parameters using a neural network similar to \cite{pulkki_MLAcoustics:2019} to learn the mapping from 3332 rooms 
%\nr{ziqi: is it really 3000 rooms? I thought they had very few room geometries and then the shelves moved around? What were those numbers? Put the data in comment here and I'll adapt.} \zf{Actually, 1 room, and 3332 configurations} 
%to acoustical parameters such as reverberation time ($T_{60}$) and sound pressure level ($L$) \hg{is this the same as $L$ in Eq. 2? Otherwise rename / remove} at a few listener locations.
%Closest to our work, 

\vspace{2pt}

Pulkki and Svensson~\cite{pulkki_MLAcoustics:2019} trained a small fully-connected neural network to learn exterior scattering from rectangular plates as predicted by the Biot–-Tolstoy–-Medwin (BTM) diffraction model~\cite{svensson:1999}. The input was a carefully designed low-dimensional representation of the geometric configuration of source, plate, and listener based on knowledge of diffraction physics. The output was a set of parameters of low-order digital filters meant to auralize the effect. The authors report plausible auralization of scattering effects despite some inaccuracies. However, due to relying on a hand-crafted low-dimensional parameterization, the method is not designed to generalize beyond rectangular plates. 

%recently learned exterior scattering from rectangular plate geometries using a fully-connected neural network with about 10 neurons each, trained on 78 plates. 
%The input to the network  were acoustically-aware features closely based on BTM  edge diffraction theory \cite{svensson:1999}, to describe the configuration of source-plate-receiver. Object shape was approximately represented using PCA vectors. 
%The output were parameters of low-order filters such as cutoff frequencies to render frequency-dependent low- and high-pass effects. Although the results weren't highly accurate the authors showed that plausible auralization could be obtained nevertheless.
%While the authors report that their approach could be used to produce plausible auralizations despite some inaccuracies \nr{inaccuracy isn't a put-down here, Ville used this characterization in his analysis of results in the paper}, it is limited to modelling the scattering effect of rectangular plates, and an extension to general shapes is not trivial (reference?) \zf{The only paper that mentions general shape is \cite{falcon2018machine}, which we already cited.}. \nr{I have re-phrased this so that it doesn't (incorrectly) seem like we're making a broad assertion without support. The technique is designed from the ground up for plates, no wonder it doesn't extend.}

\begin{figure}
    \centering
    \includegraphics{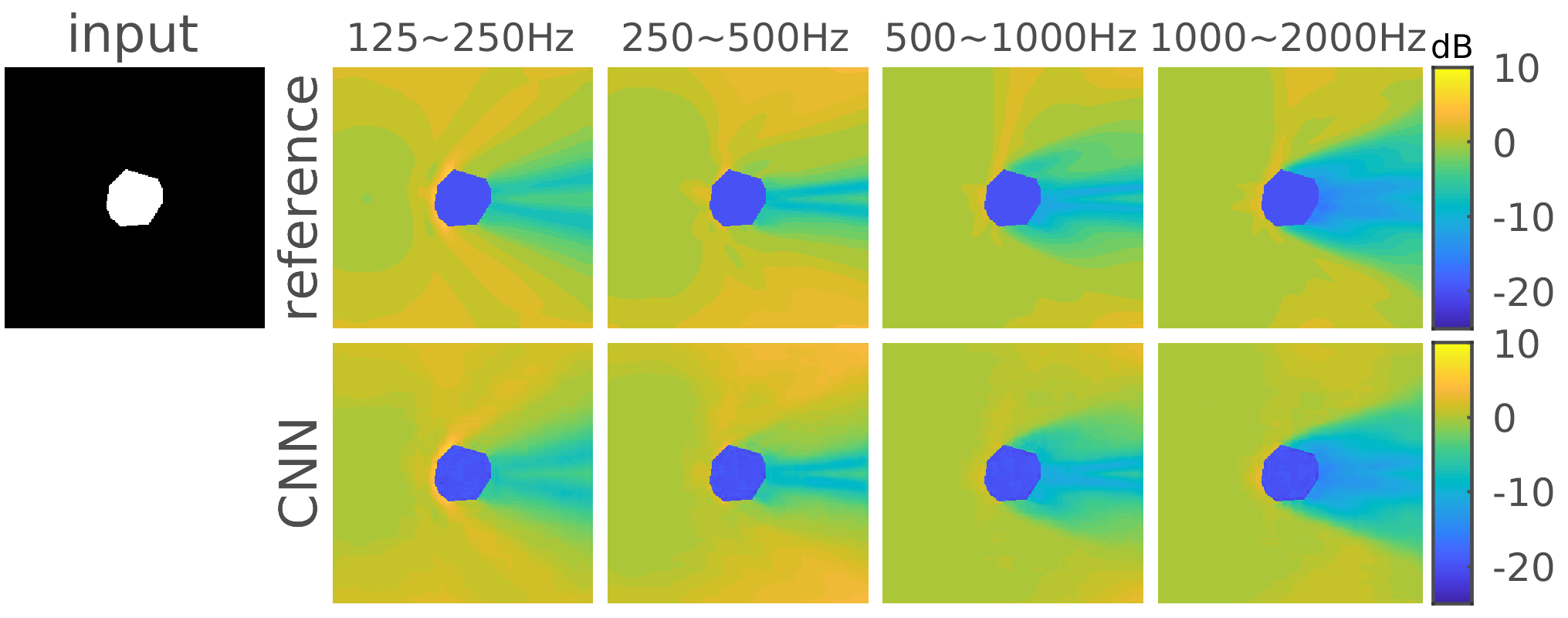}
    %\vspace{-16pt}
    \caption{Acoustic scattering formulated as 2D image-to-image regression. Input object shape is specified as a binary image (left). A point source, not shown, is placed to the left of the object. Numerical wave simulation is used to produce reference scattered loudness fields in frequency bands (top row). Our CNN  produces a close approximation at over 100$\times$ speedup (bottom row).}
    \label{fig:input-output}
\end{figure}

In this paper, we report the first study on whether a neural network can effectively learn the mapping from a large class of shapes (convex prisms) to the resulting frequency-dependent loudness field, as illustrated in \cref{fig:input-output}. We restrict the problem to convex shapes to rule out reverberation and resonance effects in this initial study. In contrast to \cite{pulkki_MLAcoustics:2019}, our goal is to design a neural network that generalizes well for a variety of input shapes by formulating the problem as high-dimensional image-to-image regression which allows application of state-of-the-art convolutional neural networks (CNNs) that have been successfully applied in computer vision \cite{he2016deep, ren2015nips, he2017iccv}. 

We design a CNN that ingests convex prism geometries represented by their 2D cross-sections discretized onto binary occupancy grids. The predicted outputs are the corresponding loudness fields in octave bands along a horizontal slice passing through the source, represented as floating point images in decibels (dB).
%using a $256\times 256$ occupancy grid with dense (6.35~cm) spatial resolution. The neural network learns to map this general shape representation to 4 octave-band loudness fields spanning 125-2000Hz at the same spatial resolution. Thus we learn a high-dimensional function mapping from $65536$ input to $65536\times 4$ output dimensions.
Our input--output mapping of acoustic scattering in terms of a spatial grid reveals spatial coherence, such as the smooth change in loudness across the geometric shadow edge. CNNs are particularly well-adapted to such tasks. Further, using CNNs allows us to train a single network unlike \cite{pulkki_MLAcoustics:2019}, where occluded and unoccluded cases had to be treated separately with distinct networks. 
%We employ a convolutional neural network to learn the spatial structure of the input shape as well as the resulting scattered field. Specifically, we show that by adapting full-resolution residual networks (FRRNs) \cite{pohlen2017full} proposed recently in the image segmentation community, spatially detailed acoustic fields can be predicted surprisingly well. We train our nerual network on \numprint{108000} random inputs and report RMS errors as small as 2 dB on a disjoint random test set (Fig. \ref{fig:comparison}). We further test generalization ability by carefully picking a few shapes well outside the training data and still find good agreement (Fig. \ref{fig:error}). Our results show that convolutional neural networks offer a surprisingly effective avenue for fast and spatially detailed prediction of acoustic scattering by carefully crafting the input-output representation to exploit their strength, and a large training set to avoid over-fitting. We also share our dataset to foster further research in this exciting area.

Experimental results and generalization tests indicate that the proposed neural network model is surprisingly effective at capturing detailed spatial variations in diffracted scattering and occlusion (e.g., compare top vs.\ bottom row in \cref{fig:input-output}). Relative to wave simulated reference, the RMS error is below 1dB while providing over 100x speedup, with evaluation time of about 50ms on a high-end GPU. To foster further research, we have shared our complete dataset at: \url{https://github.com/microsoft/AcousticScatteringData}.

\section{Problem Formulation}
\subsection{Acoustic loudness fields}
\label{sec:loudnessfield}
Consider the exterior acoustics problem of an object insonified with a point source at location $\mathbf{x}_0 = (x_0, y_0, z_0)$ emitting a Dirac impulse. 
Object shape can be abstractly described with an indicator function, $\obj(\mathbf{x})=\{0,1\}$, where 1 indicates the object is present at a 3D spatial location $\mathbf{x}$, and 0 indicates otherwise. Scattering from the object results in a time-varying pressure field denoted by $G(\mathbf{x},t;\xs)$ termed the Green's function, which evaluates the pressure at any point $\mathbf{x} = (x,y,z)$ at time $t$. Semi-colon denotes parameters to be held fixed; in this case the source location, $\xs$. The Green's function must satisfy the scalar wave equation,
\begin{equation}
    \left[\partial_t^2-c^2\nabla^2\right]G(\mathbf{x},t;\xs)=\delta (\mathbf{x}-\xs,t),
    \label{eq:wave}
\end{equation}
where $c=343$~m/s is the speed of sound and $\nabla^2=\frac{\partial^2}{\partial x^2} + \frac{\partial^2}{\partial y^2} + \frac{\partial^2}{\partial z^2}$ is the Laplacian operator, subject to the impedance boundary condition on the object's surface based on its material, and the Sommerfeld radiation condition at infinity \cite{pierce:1989}. 

Analytical solutions to \cref{eq:wave} are unavailable beyond simple geometries such as a sphere or an infinite wedge \cite{pierce:1989}. Therefore, numerical solvers must be employed 
that perform computationally expensive 
%which are computationally expensive since they must perform 
sub-wavelength sampling of space. 
For applications such as modeling dynamic occlusion in virtual reality or optimizing baffle shape in noise control, the energetic properties of $G$ are of particular interest, obtained by measuring its loudness in frequency bands.
Therefore, the focus of our study is the sensitive, non-linear effect of object shape on the scattered loudness field. 

Formally, denoting the temporal Fourier transform as $\mathcal{F}$, we define the Green's function in the frequency domain for angular frequency, $\omega$: $\hat{G}(\mathbf{x},\omega;\xs) \equiv \mathcal{F}\left[G\left(\mathbf{x},t;\xs\right)\right]$ and define octave-band loudness fields as
\begin{equation}
    L_i(\mathbf{x};\xs) \equiv 10 \log_{10} \frac{\lVert\mathbf{x}-\xs\rVert^2}{\omega_{i+1}-\omega_i} \int_{\omega_i}^{\omega_{i+1}} \lvert\hat{G}(\mathbf{x},\omega;\xs)\rvert^2 d\omega,
    \label{eq:encoding}
\end{equation}
where $i\in\{1,2,3,4\}$ denotes the index of four octave bands  $\left[\omega_i,\omega_{i+1}\right)$, $\omega_i \equiv 2\pi \times 125 \times 2^{i-1}$ rad/s, which together span the frequency range of [125, 2000] Hz. 
The factor $\lVert\mathbf{x}-\xs\rVert^2$ normalizes $L_i$ for free-space distance attenuation, so that in the absence of any geometry, $L_i(\mathbf{x};\xs)=0$. That is, all loudness fields are 0~dB everywhere in the absence of a scatterer and they capture the \emph{perturbation} on free-space energy distribution induced by the presence of object geometry, which is often the primary quantity of interest. Distance attenuation can be easily included later via a compensating factor of $1/\lVert\mathbf{x}-\xs\rVert^2$.

\subsection{Scattering functional}
\label{sec:scattering func}
From \cref{eq:wave}, the loudness fields $L_i$ depend both on the object geometry $\obj$ and source location $\xs$. We observe that the latter can be restricted to the negative x-axis, simplifying the formulation, as the D'Alembert operator, $\left[\partial_t^2-c^2\nabla^2\right]$ is invariant to the choice of frame of reference \cite{pierce:1989}. Thus, given any $\xs$ in one frame of reference with origin at object center, one can find a unique coordinate system rotation $\rot$ such that $\rot(\xs)$ lies on the negative x-axis in the new coordinate system. The object must also be rotated so that $\rot({\obj})$ and evaluations of the loudness fields similarly transformed to the rotated system. 
Therefore, the source can be restricted to the negative x-axis without any loss of generality because we are approximating scattering from \emph{arbitrary} convex shapes and rotation preserves convexity.

%In short, as long as one studies scattering from \emph{arbitrary} convex shapes (which implicitly include rotated versions of a shape) the source can be restricted to the negative x-axis without any loss of generality.

The remaining free parameter for the source is its radial distance to object center.
In this initial study, distance is assumed to be fixed. This simplification allows dropping the dependence on $\xs$ entirely. The problem can then be formalized as computing the \emph{scattering functional}, $\mathcal{S}:\obj\mapsto \{L_i\}$ which takes object shape as input and outputs a set of loudness fields in frequency bands. The functional is typically evaluated using a numerical solver for \cref{eq:wave} coupled with an encoder that implements \cref{eq:encoding}, such as in the ``Triton'' system \cite{Raghuvanshi:2018:Triton} that we employ as baseline. The underlying solver has been validated in outdoor scenes \cite{Mehra_ARD_OutdoorValidation:2014}. Here we investigate whether neural networks may be used to provide a substantially faster approximation of $\mathcal{S}$.

\begin{figure}
    \centering
    \includegraphics{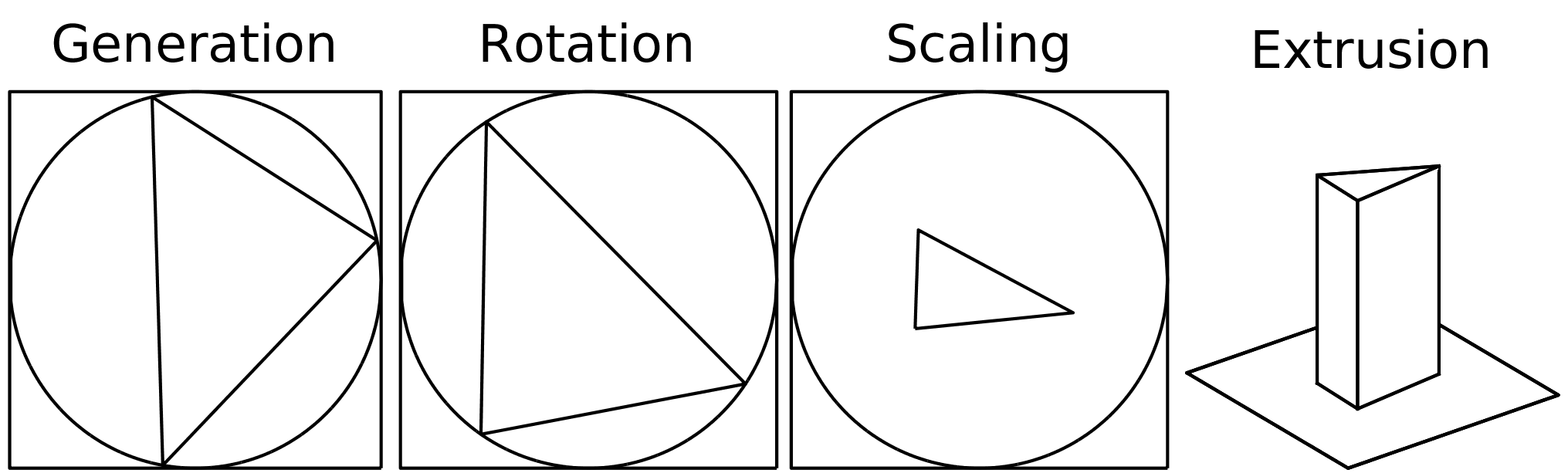}
    \caption{Random object generating process. A polygon with 3-20 vertices is randomly generated by sampling angles on a circle, then rotated, scaled, and extruded in height to yield a convex prism object. }
    \vspace{-3pt}
    \label{fig:shape generation}
\end{figure}

\begin{figure}
    \centering
    \includegraphics{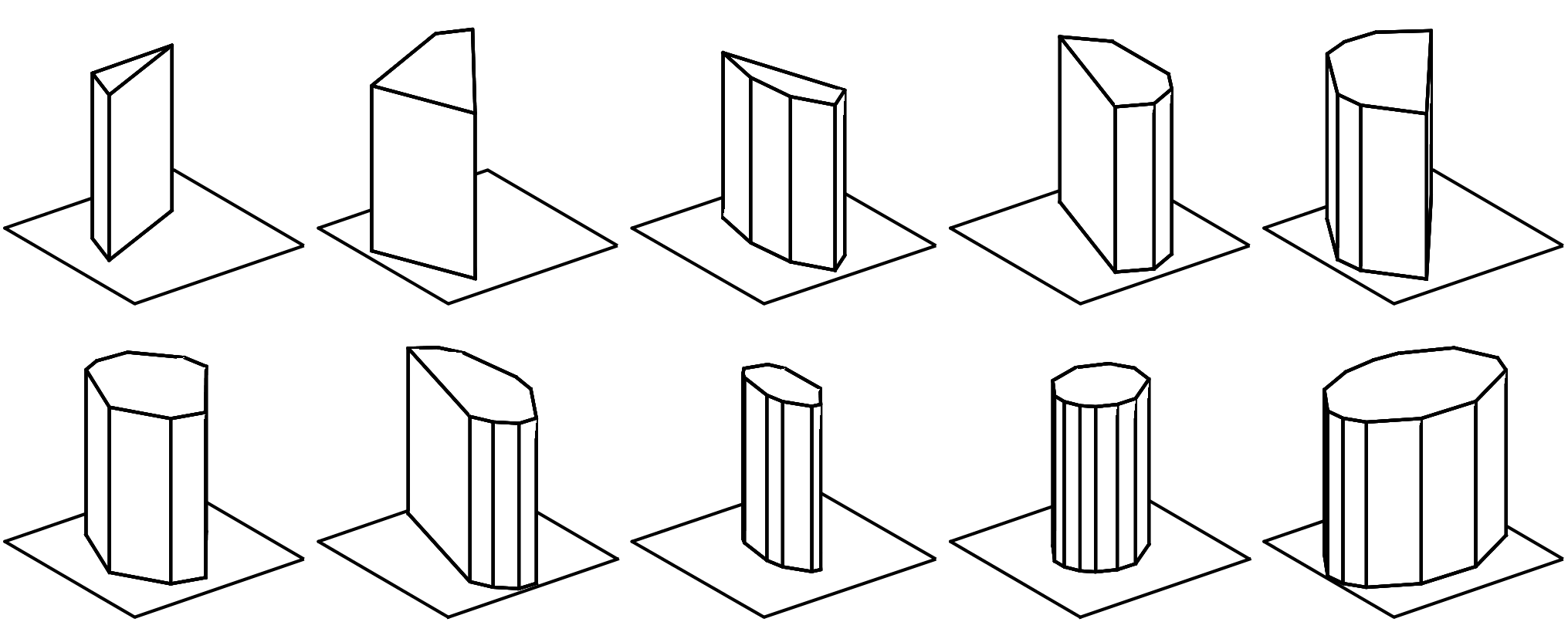}
    \caption{Examples objects in our training dataset.}
    \vspace{-3pt}
    \label{fig:diversity of shapes}
\end{figure}

\subsection{Acoustic scattering as image-to-image regression}
\label{sec:formulation2}

In order to learn $\mathcal{S}$ successfully using a neural network, the choice of discrete representation for input $\obj$, output $L_i$, and neural network architecture are critical inter-dependent considerations. We observe that shapes and loudness fields exhibit joint spatial coherence, containing smoothly varying regions, occasionally interrupted by abrupt changes such as near the object's edges, or near the geometric shadow boundary.   
%\hg{they are spatially smooth - though arguably a very important feature here and in images are edges - so perhaps ``strong local spatial coherence''?} \nr{OK read again, how does current writeup above sound? If decent, remove both comments.} 
Convolutional neural networks (CNNs) have been used extensively in the computer vision community for signals with such piece-wise smooth characteristics, motivating our current investigation. However, CNNs typically work on images represented as 2D grids of values. Therefore, we cast our input--output representation to 2D by restricting our shapes to convex prisms that have a uniform cross-section in height, i.e., along the z-axis, and training the neural network to map from this 2D convex cross-section to a 2D slice of the resulting 3D loudness fields.
The simulation setup is shown in \cref{fig:simulation region} and detailed in Section \ref{sec:simulation setup}. Thus, the task is simplified to that of image-to-image regression, as illustrated in \cref{fig:input-output}. The input is a binary image specifying presence of object, $\obj$, at each pixel, and output is a multi-channel image with four channels corresponding to the four octave-band loudness fields $L_i$.

\section{Data generation}
\label{sec:data generation}
The data generation consists of generating random convex-prism input shapes and computing the corresponding output loudness fields.

%For practical data generation, we employed Triton, a parametric wave field framework \cite{Raghuvanshi:2018:Triton} that supports fast incremental processing of impulse responses. We also extended the framework of Triton to compute the loudness fields $L_i$ using standard recursive octave-band digital filters. In the end, we computed \cref{eq:wave} and \cref{eq:encoding} for more than \numprint{100000} objects and used the results as our input-output pairs.

\subsection{Input shape generation}
\label{sec:generation of objects}
%We generate a large number of random convex prisms. The footprint for all generated shapes must lie in a $4\times4$~m$^2$ ``object region.'' Shapes are generated as follows. 
The generation of random convex prisms is illustrated in \cref{fig:shape generation}.
Given a target number of vertices, $\mathrm{N}$, of the convex cross-section, 
the angles $\theta_i, i = [1, \cdots, \mathrm{N}]$ are drawn randomly from $[0,2\pi]$ and then sorted. 
%which are sorted in increasing order to obtain ordered angular samples, $\theta_i$. 
The ordered set of points $(x, y) = (2\cos\theta_i,2\sin\theta_i)$ describes a convex polygon 
%\zf{modified to $2\cos\theta_i$, as the radius is 2, and diameter is 4} and consecutive vertices 
%are connected by edges to obtain 
with all its vertices on the inscribed circle of a $4\times4$~m$^2$ \emph{object region}.
A random rotation in $[0,2\pi)$ is performed about the origin, 
%Restricting vertices to inscribed circle ensures that the generated shape is convex and rotation retains containment in the object region. 
followed by scaling in x and y independently with scaling factors drawn randomly from $[0.25,1]$. 
Finally, the rotated and scaled convex polygon is extruded along the z-axis to obtain a convex prism. All random numbers are drawn from the uniform distribution.
The procedure results in objects with significant cross-section diversity, see \cref{fig:diversity of shapes}.
%To train and evaluate the proposed neural network, separate training, validation, and test sets were generated.
%We create three separate input-output datasets: a training set, a validation set, and a test set.
For each cross-section vertex count, 
%Each cross-section vertex count, 
$\mathrm{N}\in[3,20]$, $\mathrm{K}$ random convex prisms are generated,
where 
$\mathrm{K_{tr}}=$~\numprint{6000} for the training set, 
$\mathrm{K_{cv}}=$~\numprint{60} for the validation set, 
$\mathrm{K_{te}}=$~\numprint{20} for the test set,
%For each $n$, we respectively generated \numprint{6000}, \numprint{60} and \numprint{20} objects as described above, 
resulting in a total of \numprint{108000}, \numprint{1080} and \numprint{360} samples for training, validation and test, respectively.
%\hg{add 1-2 sentences about how the binary input image is generated, then say these images were compared to ensure zero overlap between training and test sets.} 

%The test set was compared against the training set to ensure that there was no overlap in the sense that discretizing all objects onto input binary image, no binary image in the test set matches any binary image in the training set.

\begin{figure}
    \centering
    \includegraphics[width=\columnwidth]{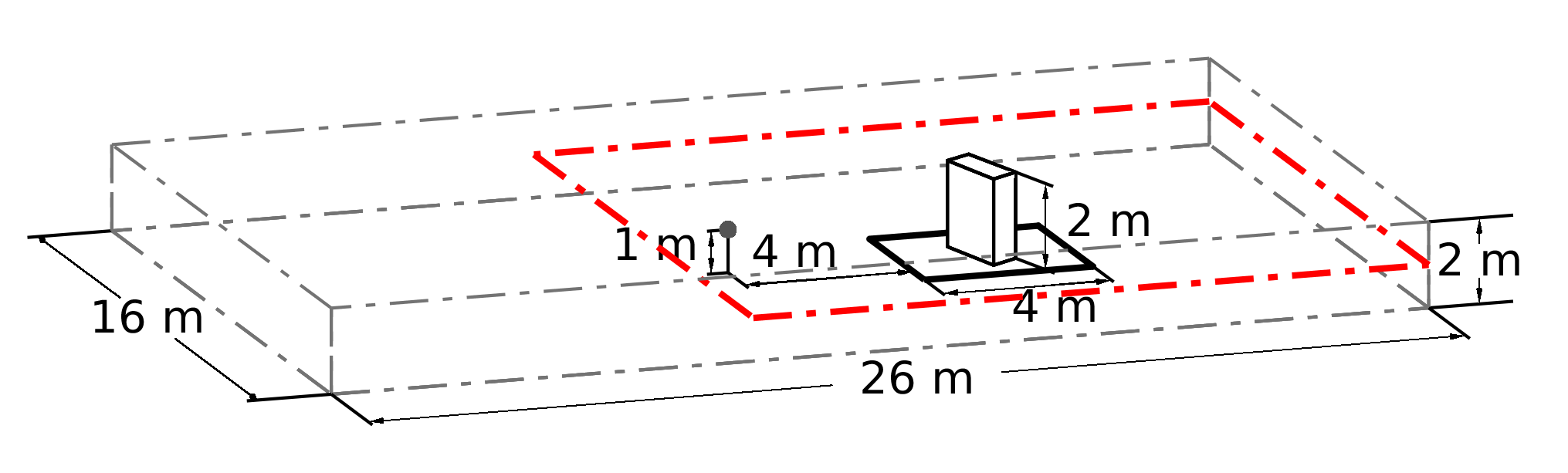}
    \caption{A convex-prism object is insonified with a point source marked with gray dot. Simulation is performed inside a containing cuboidal region. Object and loudness field data is extracted on a 2D slice shown with dashed red square. Dimensions not to scale. %\hg{Perhaps mark the source with $\times$ so you can refer to it in the caption as ``Source ($\times$) ...''. Alternatively, say ``Source, marked by a grey/gray dot, ...''}
    }
    \vspace{-12pt}
    \label{fig:simulation region}
\end{figure}

\subsection{Output loudness field generation}
\label{sec:simulation setup}
For each convex prism object, we compute the corresponding output loudness fields using the Triton system \cite{Raghuvanshi:2018:Triton} that employs the fast ARD pseudo-spectral wave solver \cite{raghuvanshi_ARD:09} to solve \cref{eq:wave} combined with a streaming encoder to evaluate \cref{eq:encoding}. The scattering object resides in a $4\times 4\times 2$~m$^3$ object region. The center of this object region is the origin of our coordinate system. We assume a nearly-rigid and frequency-independent acoustic impedance corresponding to Concrete material, with pressure reflectivity of 0.95. High reflectivity is chosen to ensure there is substantial reflection from the object.

The simulation is performed on a larger $26\times 16\times 2$~m$^3$ cuboidal region of space, as illustrated in \cref{fig:simulation region}, with perfectly matched layers absorbing any wavefronts exiting this region. A point sound source is placed on the negative x-axis at $(-6,0,0)$~m. The solver is configured for a usable bandwidth up to \numprint{2000} Hz, resulting in an update rate of \numprint{11765} Hz. The solver executes on a discrete spatial grid with uniform spacing of $6.375$~cm in all dimensions.%, ensuring above-Nyquist sampling for the shortest propagating wavelength of $17$~cm at \numprint{2000} Hz.

For extracting the input-output data for training purposes our region of interest is the $16\times 16$~m$^2$ 2D slice that symmetrically contains the object region, with corners $(-8,-8,0)$ to $(+8,+8,0)$~m, shown with red square in \cref{fig:simulation region}. The solver already discretizes the object and fields onto a 3D spatial grid for simulation purposes, so we merely extract the relevant samples lying on our 2D slice of interest from the 3D arrays, without requiring any interpolation. The extracted 2D arrays are then padded to 256$\times$256 pixel images. This results in a pair of an input binary image for an object and an output set of four loudness fields, constituting one entry in our dataset. 

We ensure the training and test sets are disjoint by exhaustively checking that none of the object binary images in the test set have an exact match in the training set. Dataset generation was run in parallel for all shapes on a high-performance cluster, taking \numprint{3} days. Each entry took 4 minutes for simulation and encoding, excluding task preparation time. An example is shown in \cref{fig:input-output}, top row. %Energy reflected back towards source is clearly visible as bright yellow region on the left side of the object and occlusion is seen as dark blue lobes to the right side of the object. The shadowing gets more pronounced with increasing frequency, as expected from diffraction physics. Casting acoustical features into such visual features is key to applying CNNs to the problem.

\section{Full-Resolution Residual Network (FRRN)}
\begin{comment}
\begin{figure}[b!]
    \centering
    \includegraphics[width=0.95\columnwidth]{Images/FRRN.png}
    \caption{An illustration of the full-resolution residual network (FRRN, upper) and the full-resolution residual unit (FRRU, lower).}
    \label{fig:my_label}
\end{figure}
\end{comment}
\input{text/atchitecture.tex}

We adopt the full-resolution residual network (FRRN) \cite{pohlen2017full} to model the scattering functional defined in \cref{sec:scattering func} using the training data generated in the previous section. As shown in \cref{fig:frrn}, an FRRN is composed of two basic streams: a pooling stream and a residual stream. In general, data abstraction with multiple resolutions in the pooling stream enables the FRRN to integrate both fine local details and general transitions of loudness fields. The residual stream of full resolution ensures that loudness fields are output at the input spatial resolution and that backpropagation converges faster \cite{he2016deep,he2016identity}. 
%\zf{label residual and pooling streams; local and far range to replace...,faster convergence rate}

The core component of an FRRN is the full-resolution residual unit (FRRU), shown in \cref{fig:frru}. There are 27 FRRUs in our FRRN (3 in each FRRU block in \cref{fig:frrn}), which is the depth of our neural network. In each FRRU, the full-resolution input residual stream $\mathbf{z}_n$ is down-sampled to the same resolution as the input pooling stream $\mathbf{y}_n$ and is then concatenated to $\mathbf{y}_n$. The concatenation is fed into two consecutive convolutional units to generate the output pooling stream $\mathbf{y}_{n+1}$, which serves as the input pooling stream of the next FRRU. Further, the  stream $\mathbf{y}_{n+1}$ propagates into another convolutional unit and is upsampled to the same resolution as $\mathbf{z}_n$. The upsampled stream is added back to $\mathbf{z}_n$ to form the output residual stream $\mathbf{z}_{n+1}$, which is subsequently added back to the main stream of full resolution. Such bidirectional downsampling and upsampling of features between the residual and pooling streams allows to learn features at successive layers of FRRN at different spatial resolutions.

\section{Experimental evaluation and discussion}
\label{sec:experiments}

\begin{figure}
    \centering
    \includegraphics{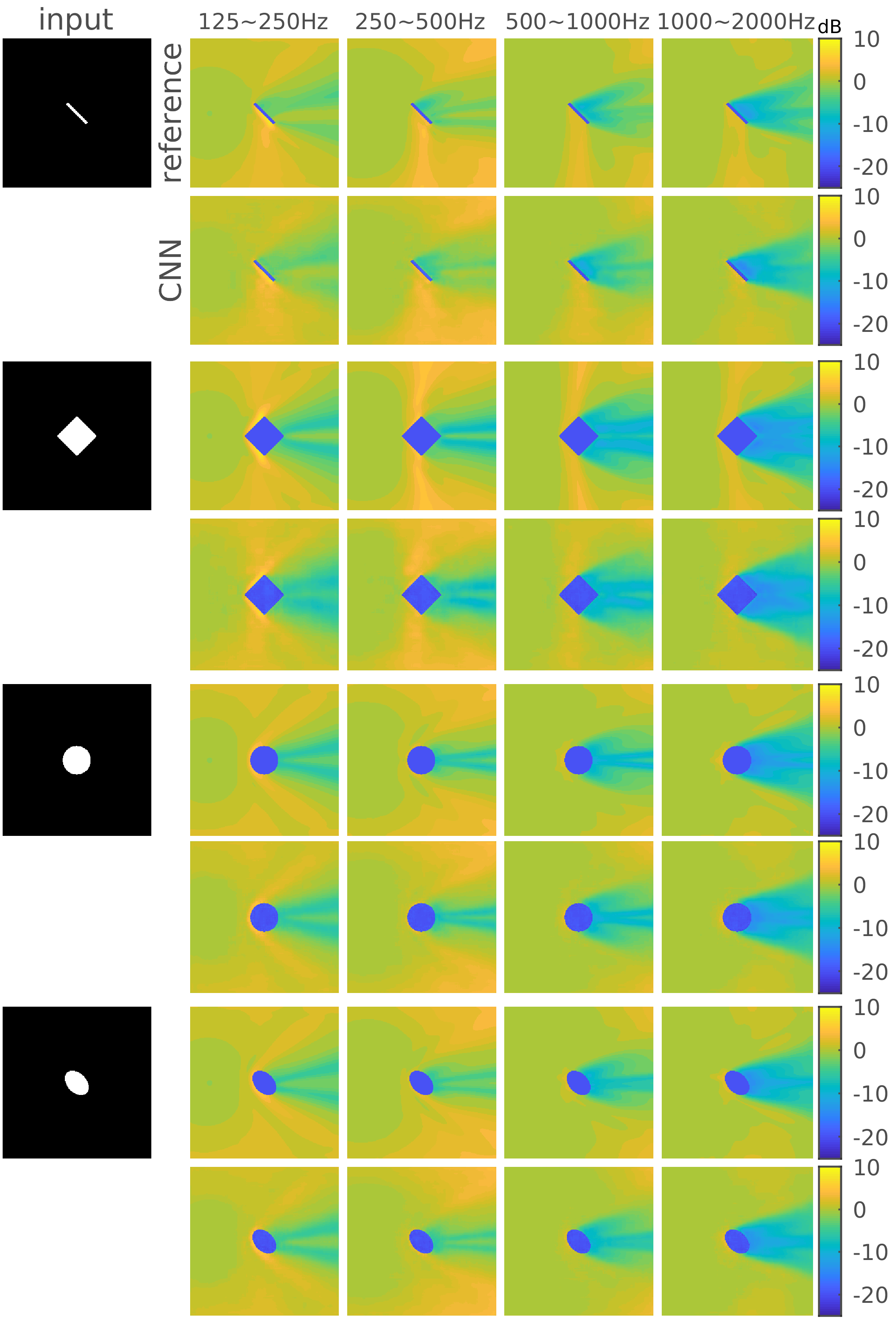}
    \caption{Generalization tests comparing reference vs CNN prediction. CNN is able to model detailed scattering and occlusion variations.}
    \vspace{-6pt}
    \label{fig:comparison}
\end{figure}

We employ the source code of FRRN provided by \cite{mshahsemseg} in our study. Since the FRRN from \cite{mshahsemseg} was originally designed for classifying pixels of images into multiple categories and modeling the scattering functional is a regression problem, we modified the source code and selected the mean squared error (MSE) as our loss function. We also modified the implementation so that the input and output of the neural network are respectively one-channel and four-channel $256 \times 256$ images, indicated by \cref{fig:input-output}. We set the batch size as 8 and adopted a stochastic-gradient-descent (SGD) optimizer, a learning rate of $1.0\mathrm{e}{-4}$, a momentum of $0.99$ and a weight decay of $5.0\mathrm{e}{-4}$. The FRRN was trained on \numprint{108000} examples for \numprint{50000} iterations on a Tesla P100 GPU. Evaluating the CNN after training takes about $50$~ms. The wave simulation takes 4 minutes on a multi-core CPU and can be accelerated by 10$\times$ if also performed on the GPU \cite{mehra:2012}. Adjusting for hardware differences, then our method is 100-1000$\times$ faster.

\begin{figure}
    \centering
    \includegraphics{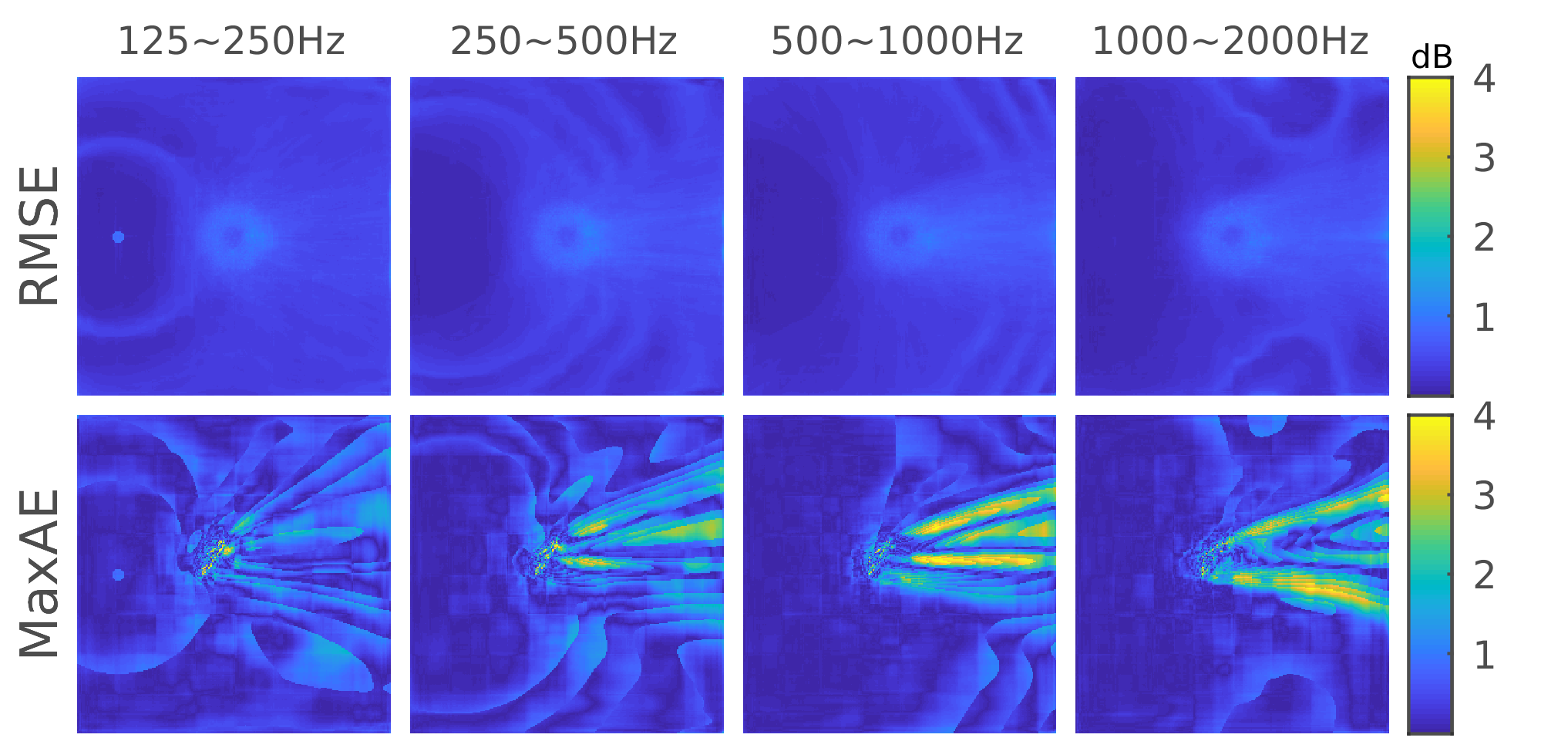}
    \caption{Root-mean-squared error (RMSE) and maximum absolute error (MaxAE) computed over 360 test cases for each frequency band.}
    \vspace{-6pt}
    \label{fig:error}
\end{figure}

To test the generalization capability of our model, we created four prisms extruded from a bar, square, circle and ellipse, all of which cause the network to extrapolate beyond the randomly-generated training set. The CNN provides a surprisingly good reproduction of the spatial loudness variation, as shown in  \cref{fig:comparison}. Notice the reflected lobe from the bar prism (top row), which shows scattered energy propagating downwards. In comparison, reflection from the square prism (second row) is symmetric about the x-axis. The CNN successfully predicts these different acoustic features, even though it has only learned from random polygons. The CNN does introduce a degree of spatial smoothing on the scattered lobes, a trend we observe consistently. Also note the brightening at low frequencies at edges facing the source. This is due to constructive interference between incident and reflected signals. %The CNN is able to capture such details.
%, in large part because of our use of a full-resolution network.
The CNN is also able to capture diffracted shadowing behind the object in all cases, along with smooth variation around the geometric shadow that gets more abrupt as frequency increases. Our results indicate that learning spatial \emph{fields} of perceptual acoustic quantities is quite advantageous compared to learning acoustic responses at a few points, since fields provide the network with extensive information about the constraints on spatial variation imposed by wave physics.

As a statistical test of accuracy, we fed the 360 objects in the test set into the trained network and evaluated the root-mean-squared errors (RMSE) and the maximum absolute errors (MaxAE) on all pixels in all frequency bands against the reference simulated results. These are shown in \cref{fig:error}. The RMS errors are below \numprint{1} dB for all frequency bands. % and demonstrate symmetry with respect to the x-axis. The errors further averaged over all pixels in the four octave bands are respectively 0.32 dB, 0.34 dB, 0.32 dB and 0.29 dB. 
MaxAE provides a more detailed look at errors within particular test cases. At each pixel it shows the largest absolute error over all \numprint{360} test cases. As illustrated in \cref{fig:error}, the errors are concentrated in the occluded region behind the object. This phenomenon can be explained as follows. Observe in \cref{fig:comparison} that our CNN is able to successfully predict the spatially-detailed light and dark streaks in the occluded region. These streaks are interference fringes due to diffracted wave-fronts wrapping around the object and meeting behind. Fringes oscillate faster in space for smaller wavelengths so that slight displacements in the fringes can cause large per-pixel errors due to subtracting two oscillatory functions with a small relative translation. This explanation fits the observation that MaxAE has a worsening trend with increasing frequency. Even so, our pessimistic MaxAE estimate is of the order of $4$~dB, which, while larger than the best-case just-noticeable-difference of $1$~dB, is  sufficient for plausible auralization with spatially smooth effects.

%\zf{not quite confident about this writing. Should I strengthen the writing for the comparison between simulation and prediction? The figure is intuitively straightforward, but language description is not strong enough.}

%\nr{Note that our error is in the form of spatial smoothing. Such type of error is well tolerated in auralization applications, where smooth changes on listener/source motion are very important. Such nice errors are because we use a CNN, with a densely connected net, no such guarantee, right?}
%\zf{I think the neural network indeed has the effect of smoothing, but I am not sure whether our network can be counted as densely connected, as one purpose of CNN is to reduce connection.}\nr{yes thats what I meant. A fully connected net is probably bad because it is not restricted to obey spatial locality whereas a CNN is.} 

\section{Conclusion and outlook}
\label{sec:conclusion}
We investigated the application of convolutional neural networks (CNNs) to the problem of acoustic scattering from arbitrary convex prism shapes. By formulating the problem as 2D image-to-image regression and employing full-resolution residual networks we show that surprisingly detailed predictions can be obtained. Generalization tests indicate that the network hasn't just memorized. Network evaluation is over 100$\times$ faster than direct simulation. Our results suggest that CNNs are a promising avenue for the tough problem of fast acoustic occlusion and scattering, meriting further study. 

This initial study had several restrictions: convex prism shapes only, fixed object material, fixed source distance, and training on 2D slices. Our formulation is designed so it generalizes beyond these restrictions. A natural extension of the current approach could be to employ 3D CNNs~\cite{DaiCSHFN_cvpr17} for handling arbitrary shapes and corresponding 3D loudness fields.
The limitation of fixed source distance could be addressed by providing an additional floating point input  to the neural network that parameterizes the input--output mapping.
%For instance, for handling arbitrary shapes in 3D, one would need  CNNs on 3D grids, already being studied \cite{DaiCSHFN_cvpr17, MaturanaS_iros15} \nr{Vibhav - a good 3D CNN citation?}. The object would be represented as a 3D occupancy grid resulting in 3D loudness fields. Source distance could be included as an additional floating point input, or simply as a special indicator value in the input object grid 
We intend to pursue such extensions in future work, and hope our results and dataset foster parallel investigations in this exciting direction.

\bibliographystyle{IEEEtran}
\bibliography{strings,nikunj,refs,refs2,IEEEsettings}

\end{document}